\documentclass[draft]{agujournal2018}
\usepackage{apacite, amsmath}
\usepackage{url} %this package should fix any errors with URLs in refs.
\usepackage{lineno}
\usepackage{ifetex}
\draftfalse

\journalname{JGR: Space Physics}

\begin{document}

%% ------------------------------------------------------------------------ %%
%  Title
%
% (A title should be specific, informative, and brief. Use
% abbreviations only if they are defined in the abstract. Titles that
% start with general keywords then specific terms are optimized in
% searches)
%
%% ------------------------------------------------------------------------ %%

% Example: \title{This is a test title}

\title{Another Mini Solar Maximum in the Offing: A Prediction for the Amplitude of Solar Cycle 25}

%% ------------------------------------------------------------------------ %%
%
%  AUTHORS AND AFFILIATIONS
%
%% ------------------------------------------------------------------------ %%

% Authors are individuals who have significantly contributed to the
% research and preparation of the article. Group authors are allowed, if
% each author in the group is separately identified in an appendix.)

% List authors by first name or initial followed by last name and
% separated by commas. Use \affil{} to number affiliations, and
% \thanks{} for author notes.
% Additional author notes should be indicated with \thanks{} (for
% example, for current addresses).

% Example: \authors{A. B. Author\affil{1}\thanks{Current address, Antartica}, B. C. Author\affil{2,3}, and D. E.
% Author\affil{3,4}\thanks{Also funded by Monsanto.}}

\authors{Susanta Kumar Bisoi\affil{1}, P. Janardhan\affil{2} and S. Ananthakrishnan\affil{3}}

% \affiliation{1}{First Affiliation}
% \affiliation{2}{Second Affiliation}
% \affiliation{3}{Third Affiliation}
% \affiliation{4}{Fourth Affiliation}

\affiliation{1}{Key Laboratory of Solar Activity, National Astronomical
	Observatories, Chinese Academy of Sciences, Beijing - 100100, China}

\affiliation{2}{Physical Research Laboratory, Astronomy \& Astrophysics
	Division, Navarangpura, Ahmedabad - 380009, India.}
	
\affiliation{3}{Electronics Science Department, Pune University, Pune, 
			Maharasthra - 411007, India}

%(repeat as many times as is necessary)

%% Corresponding Author:
% Corresponding author mailing address and e-mail address:

% (include name and email addresses of the corresponding author.  More
% than one corresponding author is allowed in this LaTeX file and for
% publication; but only one corresponding author is allowed in our
% editorial system.)

% Example: \correspondingauthor{First and Last Name}{email@address.edu}

\correspondingauthor{Susanta Kumar Bisoi}{susanta@nao.cas.cn}

%% Keypoints, final entry on title page.

%  List up to three key points (at least one is required)
%  Key Points summarize the main points and conclusions of the article
%  Each must be 100 characters or less with no special characters or punctuation

% Example:
% \begin{keypoints}
% \item	List up to three key points (at least one is required)
% \item	Key Points summarize the main points and conclusions of the article
% \item	Each must be 100 characters or less with no special characters or punctuation
% \end{keypoints}

\begin{keypoints}
\item Anti-solar-cycle like behaviour of unsigned solar polar magnetic fields in cycle 24
\item A reduced floor level of HMF in upcoming minimum of cycle 24
\item Prediction of a relatively stronger upcoming solar cycle 25 than cycle 24
\end{keypoints}

%% ------------------------------------------------------------------------ %%
%
%  ABSTRACT
%% ------------------------------------------------------------------------ %%

%% \begin{abstract} starts the second page

\begin{abstract}
We examine the temporal changes in both solar polar magnetic field (PMF) at latitudes 
$\ge$ $45^{\circ}$ and heliospheric magnetic field (HMF) at 1 AU during solar cycles 21--24 with 
emphasis on the recent activity changes after July 2015, the so called  
``mini solar maximum" of cycle 24. While unsigned PMF shows a solar-cycle-like variation 
in cycles 21 and 22, it shows an anti-solar-cycle-like variation in cycle 24. In addition, 
the floor level of the HMF (of 4.6 nT),  i.e. the value that the HMF returns
to at each solar minimum, is breached about two years prior to cycle 24 minimum, indicating 
a reduced HMF floor level in the upcoming cycle 24 minimum. In light of the change of unsigned 
PMF and the availability of a revised smoothed sunspot number (SSN) after July 2015, we have 
revisited the correlation of unsigned PMF and HMF at solar minimum. The correlation is used 
to estimate a new value of the HMF of 4.16$\pm$0.6 nT at the cycle 24 minimum and the amplitude 
of the upcoming cycle 25. The updated prediction is 82$\pm$8 and 133$\pm$11, on the original (V1.0) and 
revised (V2.0) SSN scales, respectively. These better and more reliable SSN values (due to the 
larger data set) imply that we will witness another mini solar maximum in the upcoming cycle 25 
which will be relatively stronger than cycle 24 and a little weaker than cycle 23, even if 
the current solar cycle minimum occurs in 2021 instead of 2020.
\end{abstract}
%% ------------------------------------------------------------------------ %%
% 
%  TEXT
%
%% ------------------------------------------------------------------------ %%

\section{Introduction}
The solar cycle activity that waxes and wanes with a period of 11 years 
modulates the heliospheric environment and has potential implications 
for changes in ``space weather". It is, therefore, extremely important 
to understand long term changes in solar cycle activity and to accurately predict 
the behaviour of upcoming solar cycles.  A number of satellites and 
space missions in the recent years including many being planned for the 
future require knowledge of future solar cycle activity in planning the missions 
properly. The current solar cycle 24 is the fourth successive cycle, since 
cycle 21, in a continuing trend of diminishing sunspot cycles and is also one of the 
weakest cycles, since cycle 14, with a peak smoothed sunspot number (SSN) of 116 
in the revised sunspot scale. The maximum of solar cycle 24 is therefore known 
as the ``mini solar maximum".  It must be clarified here that as of July 2015 a 
revised and updated list of the (Wolf) sunspot numbers has been adopted, referred to 
as SSN V2.0 \citep{CLe16,Cli16}.  Recent studies have also claimed that the Sun 
may move  into a period of very low sunspot activity comparable with the Dalton  \citep{ZPo14} 
or even the Maunder minimum \citep{ZGk15,San16}. This has caught the attention 
of researchers worldwide who have attempted to predict the amplitude of solar cycle 25 
\citep{UHa14,JaB15,CJS16,HUp16,KKR17,IiH17,PSc18,JiW18,UHa18,PeN18,GoM18,
MSR18,SaK18,BNa18}. The solar cycle 25 predictions made prior to 2016 usually 
used the unrevised SSN observations, referred to here onward as SSN V1.0, while 
the solar cycle 25 predictions made after 2016 mostly used the SSN V2.0 observations. 
The different estimates of SSN in V1.0 and V2.0 for the amplitude of cycle 25 along with the 
ratio of peak SSN of cycle 25 to cycle 24 are summarised in Table \ref{tab-SSN}. 
Recently, \cite{Pes18} reported that the solar cycle 25 predictions which used the 
SSN V1.0 observations need to be revisited as he showed that the revised SSN V2.0 
observations have different values of SSN for the solar maxima and minima compared 
to the original SSN V1.0 observations.  In our previous solar cycle prediction 
\citep{JaB15}, abbreviated henceforth as JBA15, a peak SSN of 
$\sim$62$\pm$12 was reported for the amplitude of the upcoming solar cycle 25.  
For that prediction, we had used the original SSN V1.0 observations, with data for the 
period 1975 -- mid-2014. We therefore revisit, in this paper, our earlier prediction 
in order to update the amplitude of solar cycle 25 using the revised SSN V2.0 observations 
available after July 2015.
%
%----------------------------Begin Table 1 -----------------------------------------------
 \begin{table}
 	\begin{center}
 		\caption{Estimates of the amplitude of SSN for cycle 25 as reported by 
 		different researchers}
 		\label{tab-SSN}
 		\vspace{-0.4cm}
 		\begin{tabular}{@{}lccc@{}}
 			\hline\noalign{\smallskip}
 			Authors & $SSN_{max}$  & $SSN_{max}$  & $\frac{SSN25}{SSN24}$\\
				& (V1.0) & (V2.0) & \\
 			\noalign{\smallskip}\hline\noalign{\smallskip}
 			\cite{UHa14} &  -            &  -    		& $\sim$1 \\
 			\cite{JaB15} & 62$\pm$12     &  -    		& 0.83 \\
 			\cite{CJS16} &  -            &  -    		& $~$1 \\
 			\cite{HUp16} &  -            &  -    		& $~1$ \\
 			\cite{KKR17} & 63$\pm$11.3   &  -    		& 0.84 \\
  			\cite{IiH17} &  -            &  -    		& $<1$ \\
  			\cite{KiA18} & 50-55         &  -    		& 0.73 \\
  			\cite{PSc18} &  -            & 135$\pm$25	& 1.16 \\
  			\cite{JiW18} &  -            & 125$\pm$32       & 1.08 \\
  			\cite{UHa18} &  -            & 110.6            & 0.95 \\
  			\cite{PeN18} &  -            & 130              & 1.12 \\
  			\cite{GoM18} &  -            &  -               & $\sim$1 \\
  			\cite{MSR18} &  -            & 99.6             & 0.86 \\
  			\cite{SaK18} &  -            & 154$\pm$12       & 1.32 \\
  			\cite{BNa18} &  -            & 118              & 1.01 \\
  			${\bf{This ~study}}$   & 82$\pm$8      & 133$\pm$11       & 1.00, 1.14 \\
 			\noalign{\smallskip}\hline
 		\end{tabular}
 	\end{center}
 \end{table}
%---------------------------End Table 1---------------------------------------------------- 
%

Further, we primarily used, in our earlier prediction (JBA15), a possible continuation of a 
steady declining trend observed in unsigned solar polar fields above latitudes of $\ge$ 
$45^{\circ}$ starting from $\sim$1995 until the minimum of cycle 24 to estimate a value of 
unsigned polar field and subsequently a value of heliospheric magnetic field (HMF) at the 
minimum of cycle 24.  The HMF value was then used as a precursor for predicting the peak SSN of 
cycle 25. The study used solar photospheric magnetic fields (SPF) data covering the period 
1975--mid-2014. Since then, we now have three additional years of observations (up to the 
current data set of Dec. 2017) of SPF.  Recently, \cite{IJB19} have claimed that the over 
20 year steady decline in unsigned polar fields reported in JBA15 showed 
an abrupt rise after July 2015 instead of a continuing declining trend.  In this paper, we have 
shown that this change in the declining trend of unsigned polar fields observed after July 2015 would 
affect the estimated value of unsigned polar fields at the upcoming minimum of solar cycle 24 as 
obtained in JBA15, and in turn our estimate of the amplitude of cycle 25. In addition, it has been argued 
by some authors \citep{JaS15,ZGk15,San16}, that we might be heading towards a Maunder like Grand 
minimum.  In fact,  sunspot numbers going back over the past 1000 solar cycles or $\sim$11,000 years 
in time using ${^{14}}$C records from tree rings and this data has been used to identify 27 grand or prolonged 
solar minima \citep{USK07}, implying  that appropriate conditions can exist on the sun to induce grand minima. 
\cite{CKa12} and \cite{KCh13} used a flux transport dynamo model to characterise the onset of grand minima 
 seen in this $\sim$11,000 year long data set and brought out several important insights:
 \begin{itemize}
 	\item gradual 
 	changes in meridional flow velocity lead to a gradual onset of grand minima, while abrupt changes lead to an abrupt 
 	onset.
 	\item one or two solar cycles before the onset of grand minima, the cycle period tends to become longer.
 \end{itemize}
 It may be noted  that surface meridional flows over cycle 23 have shown gradual variations \citep{HRi10}
 and cycle 24 started $\sim$1.3 years later than expected. There is also evidence of longer cycles before the 
 start of the Maunder and Sp\"{o}rer minimum \citep{MiK10}. Also, modelling studies of solar cycle 23, invoking 
 meridional flow variations over the cycle have shown that very deep minima are generally associated 
 with weak polar fields \citep{NMM11}.  Given these conclusions and the fact that the declining trend in 
 photospheric magnetic fields is still continuing \citep{SaJ19}, it is reasonable to raise the question about 
 the peak SSN of cycle 25 if the oncoming solar minimum of cycle 24 were to take place in 2021 instead of 
 2020 as expected. 

Thus, in this study, we re-estimate the amplitude of solar cycle 25. Importantly, we discuss the variations 
of unsigned polar fields and HMF after July 2014 which reveal some new findings; that are crucial in the context of 
recent changes in the Sun's global magnetic field behaviour.
%
%-------------------------- Begin Figure 1 --------------------------
\begin{figure}
	\centering
	\includegraphics[width=12.0cm, height=8.0cm]{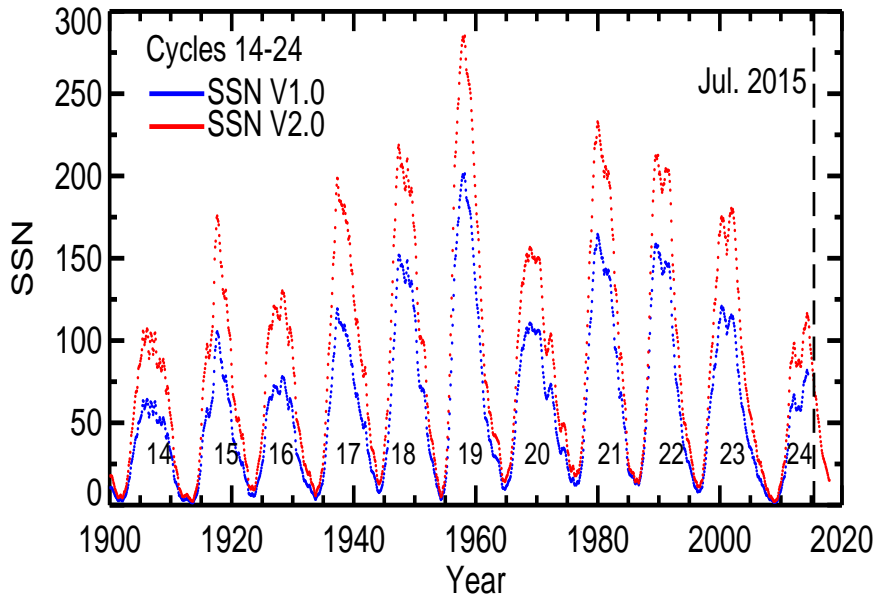}
	\caption{A comparison between the monthly SSN V1.0 (blue) and 
	the monthly SSN V2.0 (red) observations for solar cycles 14--24. 
	The recalibrated SSN V2.0 data is now Internationaly used for SSN 
	observations after July 2015.} 
	\label{fig1} 
\end{figure}
%---------------------------- End Figure 1 --------------------------
%
\section{Data and Methodology}
\subsection{Smoothed Sunspot Number (SSN)}
For SSN, we used SSN V1.0 and V2.0 observations obtained from the Royal 
Observatory of Belgium, Brussels (http://www.sidc.be/silso/datafiles). The original 
version SSN V1.0 was created in 1849 by R. Wolf who derived the daily total sunspot 
number by the formula: R$_{z}$ = Ns + 10$\times$Ng, where Ns is the number of sunspots and Ng 
is the number of sunspot groups. The original version is maintained at Zurich observatory, while a 
recalibrated SSN V2.0 version was devised after Jul. 2015 and is maintained by the 
Royal Observatory of Belgium. Observations of the SSN are available since 1749.  However, in this 
study, we have preferred to use SSN V1.0 and V2.0 observations from cycles 14--24 since only the HMF 
values of cycles 14--24 were used in this study. Figure \ref{fig1} plots observations of 
the monthly SSN V1.0 (in blue) and SSN V2.0 (in red) spanning solar cycles 14--24.  It is apparent 
from Fig.\ref{fig1} that there is no large change observed in temporal variation of SSN V2.0 except 
for the V2.0 values being about 40\%--70\% higher than V1.0. Also, there is no simple scaling factor 
between V1.0 and V2.0 that could be used for re-calculation of the predicted peak SSN of solar cycle 25.
We have therefore obtained the peak values of SSN in V1.0 and V2.0 during the solar maxima of cycles 14--24 and 
listed them in Table \ref{tab-SSNmax}. It is clear from Table \ref{tab-SSNmax} that the peak SSN 
for solar cycles 14--24 in the two versions have different values. Earlier, JBA15 directly employed 
the correlation equation proposed by \cite{CLi11}.  Based on the correlation 
between the HMF at solar minimum (B${_{min}}$) of the preceding cycle (n-1) and the peak value of 
sunspot number smoothed over a period of 13-month (SSN${_{max}}$) of the next cycle (n) these authors reported 
a correlation equation given by 

 \begin{equation}
  SSN{_{max}}= 63.4 \times B{_{min}} - 184.7
 \label{eqCL}
 \end{equation}

The peak values of SSN V1.0, from solar cycles 14 to 23, used by \cite{CLi11} was from the 
National Oceanic and Atmospheric Administration Geophysical Data Center and is listed in the fourth 
column of Table \ref{tab-SSNmax}. From a comparison of the SSN values in second and fourth columns of 
Table \ref{tab-SSNmax}, it is clear that the peak SSN V1.0 used in this study for cycles 14--17 are not 
in agreement with the peak SSN V1.0 used by \cite{CLi11}. We have thus, in this paper, updated the 
correlation between B${_{min}}$ and SSN${_{max}}$, derived by \cite{CLi11} and consequently also 
updated the prediction for the amplitude of cycle 25 that was made earlier in JBA15.
%-------------------------- Begin Table 2-----------------------------------------------------
\begin{table}
 	\begin{center}
 		\caption{Estimates of the peak values of SSN for solar cycles 14--24 in V1.0, V2.0, 
 		and the one used by \cite{CLi11}.}
 		\label{tab-SSNmax}
 		\vspace{-0.4cm}
 		\begin{tabular}{@{}lccc@{}}
 			\hline\noalign{\smallskip}
 			Solar Cycles & $SSN_{max}$  & $SSN_{max}$  &$SSN_{max}$\\
				& (V1.0) & (V2.0) & (\cite{CLi11})\\
 			\noalign{\smallskip}\hline\noalign{\smallskip}
 			Cycle 14 & 64.2            &107.1   &77.0\\
 			Cycle 15 & 105.4           &175.7 &126.5\\
 			Cycle 16 & 78.1            &130.2 &93.7\\
 			Cycle 17 & 119.2           &198.6 &143.0\\
 			Cycle 18 & 151.8           &218.7 &151.8\\
  			Cycle 19 & 201.3           &285.0 &201.3\\
  			Cycle 20 & 110.6           &156.6 &110.6\\
  			Cycle 21 & 164.5           &232.9 &164.5\\
  			Cycle 22 & 158.5           &212.5 &158.5\\
  			Cycle 23 & 120.8           &180.3 &120.8\\
  			Cycle 24 & 81.9            &116.4 & -\\
   			\noalign{\smallskip}\hline
 		\end{tabular}
 	\end{center}
 \end{table}
%---------------------- End Table 2 -----------------------------------------------------
%
\subsection{Solar Photospheric Fields (SPF)}
The SPF, for this study, were computed using medium-resolution 
line-of-sight (LOS) synoptic magnetograms from the National Solar Observatory, Kitt Peak 
(NSO/KP) and the Synoptic Optical Long-term Investigations of the Sun (NSO/SOLIS) facilities.  
Each synoptic magnetogram is available in standard FITS format and represents 
one Carrington rotation (CR) or 27.2753 day averaged SPF in 
units of Gauss. The synoptic magnetograms used here were from Feb. 1975 to Dec. 2017, 
covering CR1625--CR2197 and spanning solar cycles 21--24. We computed the unsigned 
values of SPF in the latitude range $45^{\circ}$-$78^{\circ}$, referred to here as 
polar magnetic fields (PMF). Details about the computation of SPF can be further referred 
to in \cite{JaF18}. It must be noted that researchers generally use the signed values of 
polar fields for SPF studies. However, it is to be kept in mind that in the present  
study the unsigned values of polar fields in the latitude range 
$45^{\circ}$-$78^{\circ}$ have been used with the signed values of polar fields 
being used only for comparison with the unsigned polar fields. 
The signed or unsigned values of polar fields 
in the latitude range of $45^{\circ}$-$78^{\circ}$ were estimated 
by taking into account the actual magnetic field values or the absolute of the 
actual magnetic field values, respectively. 
\subsection{Heliospheric Magnetic Field (HMF)}
We used daily measurements of HMF 
obtained from the OMNI2 data base at 1 AU (http://gsfc.nasa.gov/omniweb) covering 
the period Feb. 1975$-$Dec. 2017 that span solar cycles 21--24. 
CR averaged values of HMF were derived in order to compare and correlate them with 
CR averaged polar fields during solar cycle minima. We used CR averaged values for 
1 year intervals around solar minima of cycles 20--23 \citep{WRS09} corresponding to 
CR1642--1654, CR1771--1783, CR1905--1917, and CR2072--2084, respectively.
\section{Results}
\subsection{Photospheric Magnetic Field (PMF)}
\label{polar}

%-------------------------- Begin Figure 2-------------------------
\begin{figure}
	\centering
	\includegraphics[width=8.5cm, height=11cm]{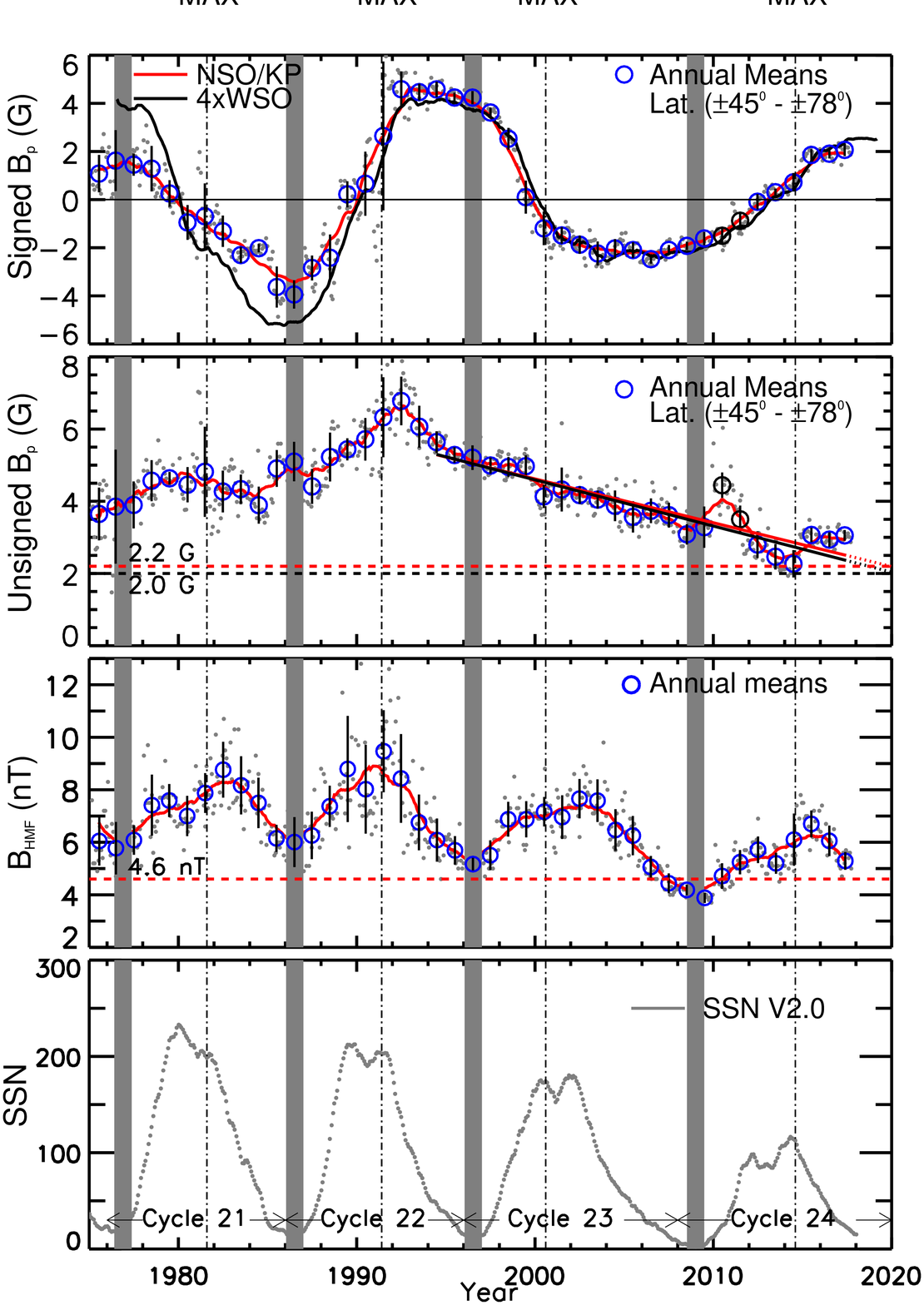}
	\caption{(first panel) variations of NSO/KP signed PMF for the time period Feb. 1975--Dec. 2017. 
	Overplotted in solid red curve are smoothed NSO/KP fields while in solid black curve is WSO signed PMF.  
	(second panel) variations of NSO/KP unsigned PMF for the same time period as in 
	the first panel.  The solid red and black lines are best fits to the declining trend for the annual means 
	while the dotted red and black lines are extrapolations of the best fits until 2020. The solid red line 
	is the best fit to all the annual means while the solid black line is the best fit to all the annual means 
	except for the years 2010 and 2011. The horizontal red and black dashed lines are marked at 2.2 G 
	and 2.0 G. (third panel) measurements of HMF during the same period. The horizontal red line 
	indicate the floor level of the HMF of 4.6 nT as proposed by \cite{SCl07}.  (fourth panel) Plotted is 
	the monthly averaged SSN V2.0 data. The filled grey dots, in the top three panels, are CR measurements of 
	SPF, while the open blue circles are annual means with 1 $\sigma$ error bars.  The vertical grey bands in 
	each panel demarcate 1 year intervals around the minima of solar cycles 20--23, while the vertical dotted 
	lines in each panel mark the solar maxima of cycles 21--24. 		
		} 
	\label{fig2} 
\end{figure}
%---------------------------- End Figure 2-------------------------
%

The first panel of Figure \ref{fig2} plots the signed PMF in the latitude range 
of $45^{\circ}$-$78^{\circ}$ for the period Feb.1975$-$Dec.2017, covering 
solar cycles 21--24. We have overplotted the smoothed NSO/KP fields (solid 
red curve) with a 13-month running mean, while for comparison, we have also 
overplotted the signed WSO PMF (solid black curve) in the latitude range poleward 
of $55^{\circ}$ for the period Apr. 1976--Apr. 2019, covering solar cycles 21--24.
It is clear that the overall temporal behaviour of NSO/KP signed PMF during solar 
cycles 21--24 show a good agreement with WSO signed PMF.  Thus, the medium 
resolution NSO/KP signed PMF are useful to study the large scale nature of 
PMF. The signed PMF in each solar cycle shows a maximum strength at the 
start of the cycle, while at solar cycle maximum, it runs through zero and changes the 
sign of the field. This is known as reversal of PMF or polar reversal. Subsequently, the 
signed PMF again shows a maximum strength during the minimum of the cycle.  
Thus, typically, the signed PMF shows an anti-solar cycle behaviour. In cycle 24, 
after the zero-crossing or reversal of PMF, the signed PMF shows a clear rise in field 
strength in the year 2015 after solar cycle maximum.  Thereafter, it shows a nearly 
constant value for the next two years in the year, {\it{i.e.}} 2016 and 2017. The 
steady value of signed PMF is evident up to Apr. 2019, about 1 year prior to the 
solar minimum of cycle 24, from the variations of WSO signed PMF.  It is to be noted that 
based on zonal and meridional flow patterns during solar cycles 23 and 24, \cite{KHH18} 
estimated that cycle 25 will begin in early 2020. Similar steady values of signed PMF 
few years before the minimum of solar cycle can also be seen during earlier solar cycles, i.e. 
cycles 22 and 23. Thus, besides the typical anti-solar cycle like behaviour of the signed 
PMF in each solar cycle it also shows steady value or a polar field plateau prior to the minimum 
of the solar cycle. 

The second panel of Fig.\ref{fig2} plots the unsigned NSO/KP PMF in the latitude range 
of $45^{\circ}$-$78^{\circ}$ for the period Feb.1975$-$Dec.2017, covering solar cycles 
21--24. It is evident from a careful examination of the strength of the unsigned PMF 
(see the solid red curve), referred to henceforth as $B{_{p}}$, that unlike the signed PMF 
the unsigned PMF shows a totally different temporal behaviour.  The unsigned PMF shows 
solar cycle like modulations in cycles 21 and 22, while, in cycle 23, no such solar cycle like 
modulation is seen. However, we see a steady decline in the field strength since the start 
of cycle 23 until the end of cycle 23. Again, at the start of solar cycle 24, in the years 2010 
and 2011, there was an increase in the $B{_{p}}$. The annual mean for the years 2010 
and 2011 is shown by open black circles. After 2011, the $B{_{p}}$ again declined from 
2012 upto 2014, only to increase again in the year 2015 after the solar maximum of cycle 24 
and has been constant since then. Thus, in contrast to the behaviour in previous cycles, the 
value of $B{_{p}}$, in cycle 24, shows an anti-solar cycle behaviour with a polar field plateau
like the signed PMF, with a maximum strength at the start of the cycle attaining a minimum 
strength around the solar cycle maximum and again showing a rise in strength after the solar 
cycle maximum and thereafter attaining a steady value. 

We further investigate whether the unexpected rise of $B{_{p}}$ during the 
year 2015 and the subsequent constant value for the next two years would 
change the declining trend.  As seen in the years 2010 and 2011, we have already witnessed 
a rise in the strength of PMF, but the declining trend of the PMF continued after 2011. Hence, 
we believe that the break in the declining trend during 2015-2017 could be temporary and that 
the declining trend may continue. However, the increase in the $B{_{p}}$ could affect its rate of 
decline, and thus, change its expected value in 2020, {\it{i.e.}} at the expected minimum of cycle 24.  
We, therefore, re-estimate the value of $B{_{p}}$ in 2020 assuming a continuing declining trend.  
The solid red line in Fig.\ref{fig2} (second panel) is a least square fit to the declining trend of $B{_{p}}$ 
for all the annual means in the period 1995--2017, while the broken red line is 
the extrapolation until 2020.  The least square fit is statistically significant 
with a Pearson correlation coefficient of $r = -0.91$, at a significance 
level of $99\%$.  Similarly, the solid black line in Fig.\ref{fig2} (second panel) 
is a least square fit to the declining trend for all the annual means, with 
the years 2010 ad 2011 being left out, and the broken black line is an 
extrapolation until 2020.  The fit is statistically significant with a Pearson 
correlation coefficient of $r = -0.94$, at a significance level of $99\%$.  The 
expected values of $B{_{p}}$ in 2020 for the above two cases are 
$\sim$2.2 ($\pm$0.08) G and $\sim$2.0 ($\pm$0.06) G, as indicated by the dashed 
horizontal lines in red and black, respectively in Fig.\ref{fig2} (second panel). 
Thus, the average expected value of $B{_{p}}$ in 2020 would be $\sim$2.1 ($\pm$0.07) G.
The expected values of $B{_{p}}$ would be $\sim$2.1 ($\pm$0.08) G and $\sim$1.9 ($\pm$0.06) G 
if the minimum is in 2021 instead. Thus, the average expected value of $B{_{p}}$ in 2021 would be 
$\sim$2.0 ($\pm$0.07) G.
 \subsection{Heliospheric Magnetic Field (HMF)}
 \label{swp}
The third panel of Fig.\ref{fig2} plots the strength of HMF (B${_{HMF}}$) at 
1 AU for the period Feb.1975$-$Dec.2017, covering cycles 21--24. 
A global reduction in the B${_{HMF}}$ from solar cycle 
22 through solar cycle 23 to solar cycle 24 is evident from 
Fig.\ref{fig2} (third panel). Also, it is seen from Fig.\ref{fig2} 
(third panel) that the HMF returns to an average value at each solar 
minimum, which is known as the floor value of HMF. The floor level 
of HMF is essentially determined by the baseline flux from the slow 
solar wind flows. \cite{SCl07} estimated the floor level of 
4.6 nT for HMF, indicated by a horizontal red line in Fig.\ref{fig2} 
(third panel), from a correlation of B${_{HMF}}$ and sunspot number 
in the period covering cycles 20--22. However, it can be seen that 
the B${_{HMF}}$, during the minimum of cycle 23, went down well 
below the floor level of 4.6 nT. Further, it is interesting to note, as seen 
from Fig.\ref{fig2} (third panel) that, in cycle 24, the B${_{HMF}}$ by the 
year 2018 has already approached the value of the floor level of 4.6 nT 
about two years prior to the minimum of cycle 24. Thus, it is expected that 
we could witness the floor level of HMF also going down below the proposed floor 
level of 4.6 nT in cycle 24. The fourth panel of Fig. \ref{fig2} plots the SSN V2.0 
observations as a function of time for the period Feb.1975$-$Dec.2017, covering 
solar cycles 21$-$24. As mentioned earlier, a global reduction of SSN since cycle 
21 upto cycle 24 is evident from Fig.\ref{fig2} (fourth panel).  
%-----------------------------------Begin Fig 3-------------------------
 \begin{center}
 	\begin{figure}
 		\includegraphics[width=14.5cm, height=9cm]{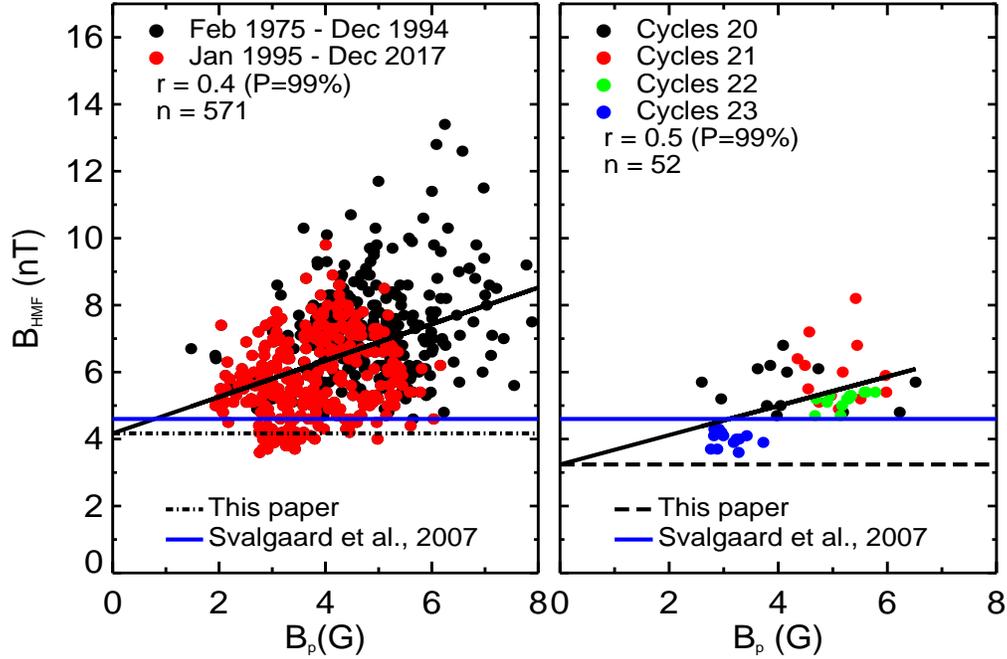} 
 		\caption{B${_{HMF}}$ as function of B${_{p}}$ shown for the period 
 		Feb. 1975--Dec. 2017 covering solar cycles 21--24 (left panel) 
		and for 1 year interval during the minima of cycles 20--23 (right panel). The 
 		correlation coefficients, r = 0.4 and 0.5, are indicated in the top right corners 
		of the panels, respectively. The solid black lines in both the panels are 
		overall fit to all data points. The filled black and red dots in the left panel are 
		measurements for the period Feb. 1975--Dec. 1994 and Jan.1995--Dec.2017, 
		respectively, while the filled dots in different colours in the right panel are 
		measurements for cycles 20--23. The 
 		solid blue lines in both panels indicate the floor levels of HMF of 4.6 nT as 
		computed by \citet{SCl07}, while the dash-dotted (left panel) and dashed (right 
		panel) horizontal lines indicate the floor levels of 4.2 and 3.2 nT, respectively, obtained in 
		this study.}
 		\label{fig3} 
 	\end{figure}
 \end{center}
%-----------------------------------End Fig 3---------------------- 
%

Since the HMF results from solar photospheric fields being swept 
out into the inner heliosphere and beyond, the declining 
B${_{p}}$ beginning around mid-1990's during solar cycle 22, and continuing during cycles 
23 and 24 would have contributed to the observed global reduction in the 
B${_{HMF}}$ and the reduction in the floor level of HMF during cycles 
23 and 24.  A plot of B${_{HMF}}$ vs B${_{p}}$ for the period 
Feb.1975--Dec.2017 is shown in Figure \ref{fig3} (left). It is evident from Fig.\ref{fig3} that 
the B${_{HMF}}$ values post-1995 (filled red-dots) show a clear reduction in their 
strength as compared to the values prior to 1995 (filled black-dots). We found a moderately 
statistically significant correlation of B${_{HMF}}$ with B${_{p}}$ with 
a Pearson correlation coefficient of $r = 0.4$ at a significance level of 
$99\%$ as indicated at the top right corner of Fig.\ref{fig3} (left).  
The solid black line in Fig.\ref{fig3} 
(left) is a best fit to all the data points between B${_{HMF}}$ 
and B${_{p}}$. The linear correlation of B${_{HMF}}$ and B${_{p}}$ can 
thus be represented by the following equation,

 \begin{equation}
 B{_{HMF}} = 4.2\pm(0.2) + (0.54\pm0.05) \times B{_{p}}
 \label{eq1}
 \end{equation}
 
which gives an intercept of 4.2 ($\pm$ 0.2) nT for B${_{HMF}}$ when 
B${_{p}}$ = 0. This implies that the floor level of the HMF would be 
4.2 nT even if solar polar field drops to zero. 
We, thus, see a reduced floor value of 4.2 nT for HMF (dotted black 
line in Fig.\ref{fig3} (left panel)), unlike the 
proposed floor level of 4.6 nT by \cite{SCl07} (solid blue line in 
Fig.\ref{fig3} (left panel)), from a correlation of B${_{HMF}}$ 
and B${_{p}}$ for the period covering solar cycles 21--24.  The 
reduced floor value of HMF is presumably due to the observed 
global reduction of the B${_{HMF}}$ post-1995. 
 
Since solar polar fields provide most of the HMF during solar minimum 
\citep{SCK05}, we now consider the correlation of B${_{HMF}}$ 
and B${_{p}}$ only during solar minima as reported in JBA15 which is 
also shown in Fig.\ref{fig3} (right panel). 
The overall fit to all data points between cycles 20--23 
is shown by a solid black line in Fig.\ref{fig3} (right) with 
an intercept of 3.2 ($\pm$0.5) nT for the HMF when B${_{p}}$ would go 
to zero and represented by the equation 

 \begin{equation}
 B{_{HMF}} = 3.2\pm(0.2) + (0.43\pm0.11) \times B{_{p}}
 \label{eq2}
 \end{equation}

The floor level of HMF is thus, $\sim$3.2 nT indicated by a dashed black 
line in Fig.\ref{fig3} (right panel), a value that has dropped by 
more than 1 nT from the proposed floor level of 4.6 nT.  
This is also due to the reduced HMF during solar cycles 23 and 24.
Thus, in order to show the contribution of the reduced HMF (due to declining 
B${_{p}}$ since the mid-1990's), we have plotted B${_{HMF}}$ Vs. B${_{p}}$ shown 
by filled dots of different colours for each solar cycle in Fig. \ref{fig3} (right panel). 
It is evident from Fig.\ref{fig3} (right panel) that the values of B${_{HMF}}$ 
are above the floor level of 4.6 nT for the minima of cycles 20--22. 
However, the values of B${_{HMF}}$ (filled blue dots in Fig.\ref{fig3} (right panel))
have gone below the floor level of 4.6 nT during the minimum of cycle 23, and 
so we see the reduction in the floor level of HMF down to 3.2 nT. As stated earlier, 
one would expect B${_{HMF}}$ to go below the floor level of 4.6 nT during the upcoming minimum 
of cycle 24 too. It means the proposed floor level could be around 3.2 nT. 
We thus preferred equation \ref{eq2}, having a better correlation than equation \ref{eq1}, 
to derive the updated expected value of B${_{HMF}}$ at the minimum of cycle 24, 
which was found to be 4.16 ($\pm$0.6) nT using the updated expected value of B${_{p}}$ 
at the minimum of cycle 24 of 2.1 G. The previous estimated value of B${_{HMF}}$ in 
2020 obtained in JBA15 was 3.9 ($\pm$0.6) nT.  If on the other hand the minimum of 
solar cycle 24 occurs in 2021, then B${_{HMF}}$ in 2021 would be 4.12 ($\pm$0.6) nT.

\subsection{Amplitude of solar cycle 25}
\begin{center}
 	\begin{figure}
 		\includegraphics[width=9cm, height=12cm]{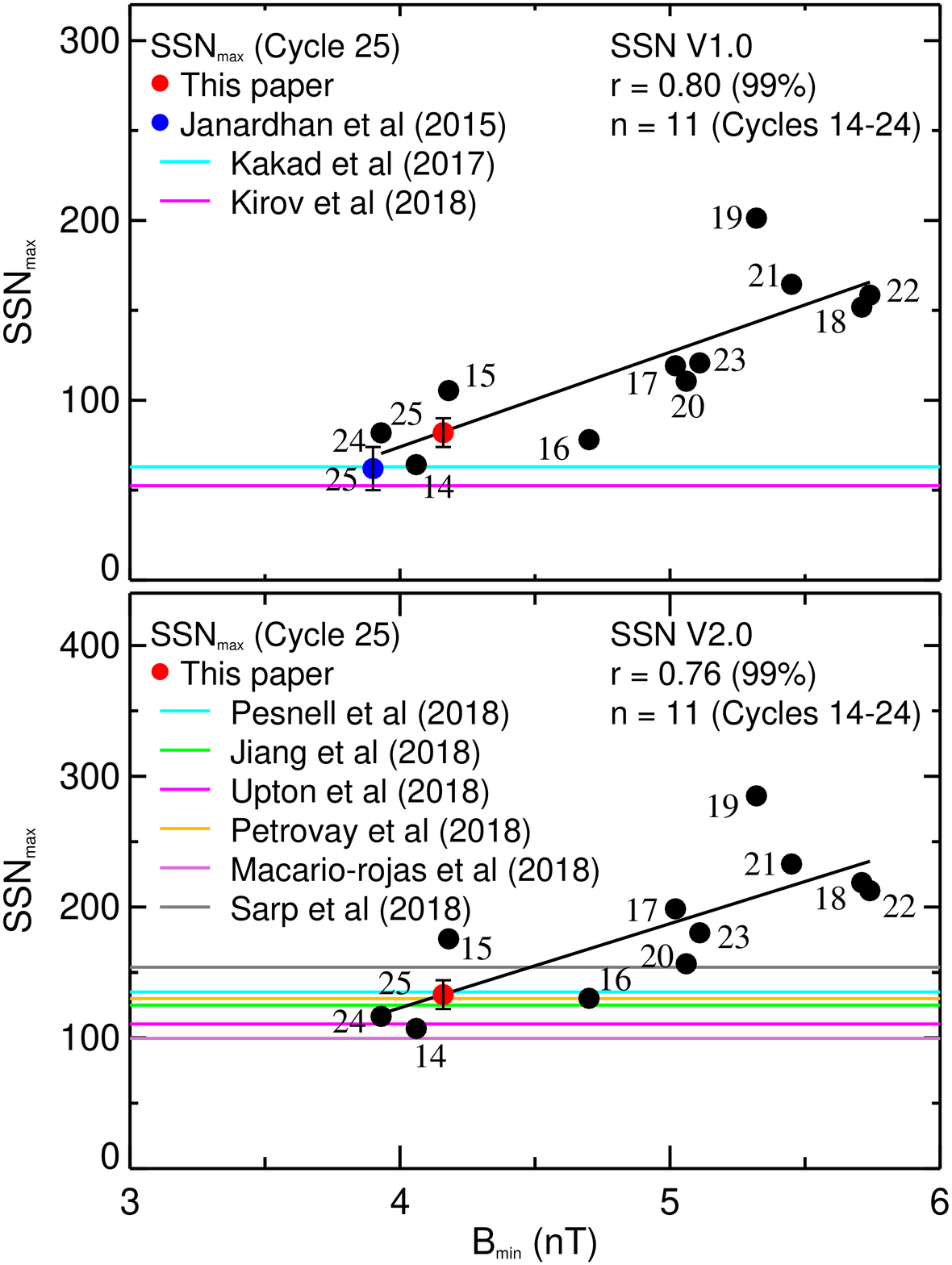} 
 		\caption{SSN${_{max}}$ of following cycles from cyles 14--24 as a 
 		function of B${_{min}}$ of preceding cycles from cycles 13-23 
 		shown for SSN V1.0 (upper panel) and SSN V2.0 (lower panel). 
 		The filled black dots with solar cycle numbers in each panel are the 
 		values of SSN${_{max}}$ as function of B${_{min}}$ for each solar cycle 
 		from cycle 14--24.  
  		The solid black line in each panel is a best fit line 
 		to all data points from cycles 14--24.
 		}
 		\label{fig4} 
 	\end{figure}
 \end{center}

The correlations between B${_{min}}$ and SSN${_{max}}$ 
using SSN V1.0 (upper panel) and SSN V2.0 
(lower panel) observations are shown in Fig.\ref{fig4}. 
The values of B${_{min}}$ used in this study are from \cite{CLi11} (see Table 
2 in \cite{CLi11}). The respective Pearson correlation coefficients, 
r = 0.80 (for V1.0) and 0.76 (for V2.0) at a significance level of 99\% are indicated at the 
top right corner of each panel. 
The correlation between B${_{min}}$ and SSN${_{max}}$ 
for SSN V1.0 is given by 

\begin{equation}
  SSN{_{max}}= 52.6 (\pm 12.9) \times B{_{min}} - 136.5 (\pm 64)
 \label{eqv1}
 \end{equation}
 
Using the value of B${_{min}}$ of 3.9 nT for cycle 24 in equation \ref{eqCL}, JBA15 
derived a SSN${_{max}}$ of 62$\pm$12 for cycle 25, indicated by a solid blue dot in the upper 
panel of Fig.\ref{fig4}. 
Using the updated value of B${_{min}}$ of 4.16 nT for cycle 24 (see section \ref{swp}) in equation 
\ref{eqv1}, we predict a SSN${_{max}}$ of 82$\pm$8 
for cycle 25. This value is indicated by a solid red dot with 1 sigma error-bar in the upper 
panel of Fig.\ref{fig4}. The two other predictions, using SSN V1.0 observations, made 
by \cite{KKR17} and \cite{KiA18} have been indicated by the coloured horizontal lines in 
the the upper panel of Fig.\ref{fig4}.
It is, thus, seen from the upper panel of Fig.\ref{fig4} that the prediction made in this study, 
using the SSN V1.0 observations that is different from the SSN observations used in JBA15, 
show clearly similar or a relatively stronger cycle 25 than cycle 24 which had a SSN${_{max}}$ of 81.9 
whereas the predictions made by JBA15 and other researchers \citep{KKR17,KiA18} indicated 
a relatively weaker cycle 25 than cycle 24.  

On the other hand, the correlation between B${_{min}}$ and SSN${_{max}}$ for SSN V2.0 data 
is given by 

\begin{equation}
  SSN{_{max}}= 64.4 (\pm 17.9) \times B{_{min}} - 134.8 (\pm 89)
 \label{eqv2}
 \end{equation}
 
Using the value of B${_{min}}$ of 4.16 nT for cycle 24 in equation \ref{eqv2}, we predict a 
SSN${_{max}}$ of 133$\pm$11 for cycle 25 if the solar minimum is in 2020.  This 
is shown by a solid red dot with 1 sigma error bar in Fig.\ref{fig4} (lower panel) with the 
differently coloured horizontal lines indicating predictions for cycle 25 by other researchers, 
using SSN V2.0 observations. The ratio of the values of SSN${_{max}}$ predicted for cycle 25 to 
the values of SSN${_{max}}$ for cycle 24 by these authors have been given in Table 
\ref{tab-SSN}. It is already clear from both Table \ref{tab-SSN} and from Fig.\ref{fig4} 
(lower panel) as to why \cite{UHa18} and \cite{MSR18} argued for a solar cycle 25 that would be 
more or less like solar cycle 24. In contrast, the prediction reported in this study using 
SSN V2.0 observations with a SSN${_{max}}$ of 133$\pm$11 for cycle 25 suggests a relatively 
stronger cycle 25 than cycle 24, which had a SSN${_{max}}$ of 116. Our prediction, in fact, 
agrees with the predictions made by \cite{JiW18}, 
\cite{PSc18} and \cite{PeN18} who claimed a similar result for the amplitude of cycle 25.
The SSN${_{max}}$ for cycle 25 would be 82$\pm$8 and 133$\pm$11 in SSN V1.0  and SSN V2.0, 
respectively, if solar minimum happens in 2020 and 80$\pm$8 and 130$\pm$11 in SSN V1.0  and 
SSN V2.0, respectively, if solar minimum happens in 2021, suggesting again a relatively stronger cycle 
25 than cycle 24, which is independent of the time of occurrence of solar minimum in cycle 24.
%-----------------------------------Begin Fig 4-------------------------
 %-----------------------------------End Fig 4----------------------
 \section{Discussion and conclusions} 
 Our study reports the temporal changes in solar photospheric fields obtained 
 from NSO/KP and NSO/SOLIS synoptic magnetograms, covering solar cycles 21--24, 
 specifically paying attention to the manner in which the unsigned 
 solar polar magnetic fields at latitudes $\ge$ $45^{\circ}$ behaved after the 
 mini-solar maximum of cycle 24. In the present study, it has been shown that after 
 the solar cycle maximum of cycle 24, unexpectedly, there has been an increase 
 in the unsigned solar polar field strength in the year 2015 that has subsequently 
 shown a slow and steady change maintaining the declining trend, that had begun around 
the mid-1990's, for more than about 22-years. Importantly, it appears 
 that the unsigned solar polar fields have switched their behaviour from 
 solar-cycle-like in cycles 21 and 22 to an anti-solar-cycle-like in cycle 24 
 and showing no correlation with solar cycle in between suggesting a transition 
 phase of solar magnetic fields during cycle 23.
 
 As per the current understanding of 
 the solar dynamo process that gives rise to the solar cycle and as proposed by 
 solar dynamo models \citep{Cha10}, sunspot or toroidal fields are generated from 
 poloidal fields by solar differential rotation, while poloidal or polar fields 
 are regenerated from toroidal fields by the Babcock-Leighton mechanism 
 \citep{Bab61,Lei69}. This mechanism depends on the systematic tilt angle 
 distribution of bipolar sunspot regions, which, in turn, is determined by the 
 Coriolis force acting on the magnetic flux tubes that rise through the solar 
 surface at different latitudes to produce bipolar sunspot regions \citep{DCh93}. 
 This whole process would result in a large scatter in tilt angle distribution, 
 which along with diffusion and surface flux transport processes could 
 change the solar polar field behaviour in any given solar cycle, therby making it totally 
 different from the previous cycle.  The sudden transition of behaviour of 
 unsigned solar polar fields in cycle 24 can, thus, be 
 attributed to the fluctuations in Babcock-Leighton mechanism that decide the net 
 flux transported towards the poles, and thereby, ultimately determining the net 
 polar field strength at the end of a cycle. 
 
 Based on the unexpected rise in solar polar fields after July 2014, we 
 have re-estimated, in this study, the new average field strength of $\sim$2.1 G 
 for the unsigned solar polar fields in 2020 for the upcoming solar minimum 
 of cycle 24. This value is quite different from the estimate of unsigned solar 
 polar fields of $\sim$1.6 G in 2020 that was reported in JBA15.   
 A reduction in the strength of HMF at 1 AU is also clearly seen post 
 July 2014, during solar cycle 24. Also, we have shown that the strength of HMF in cycle 24, post solar 
 maximum and about 2 years prior to the minimum, has nearly approached the average floor level of HMF of 
 4.6 nT as proposed by \cite{SCl07}. Using the correlation between the strength of HMF and the unsigned 
 solar polar field strength during cycles 21--24, we found a floor level of 4.2 nT. This decrease in the 
 floor level of the HMF is actually due to the observed global reduction in the HMF during the last two 
 solar cycles 23 and 24. 
 Using the correlation 
 between the strength of HMF and the unsigned  solar polar photospheric field 
 strength at minima of cycles 20-23, 
 we have estimated a value of 4.16 for the HMF in 2020 for the upcoming minimum of cycle 24.
 Further, based on the correlation of HMF  at solar minimum and SSN${_{max}}$ from 
 cycles 14--24, we estimated a value of SSN${_{max}}$ of 82$\pm$8 (V1.0) and 133 $\pm$11 
 (V2.0) for the upcoming solar cycle 25. Also, expecting a delay in the minimum of cycle 24 
 by one year, {\it{i.e.}} 2021, we also estimated a value of SSN${_{max}}$ of 80$\pm$8 (V1.0) and 130$\pm$11 
 (V2.0) for the upcoming solar cycle 25 if the minimum of cycle 24 would occur in 2021 instead of 2020.
 Our estimate suggests that the oncoming sunspot 
 cycle 25 will be relatively stronger than cycle 24, but will be weaker 
 than cycles 23.  This is different from the prediction made by JBA15 that reported rather 
 a relatively weaker cycle 25 than cycle 24. 
 
 Using the values of solar spectral 
 irradiance at 10.7 cm (F10.7) and the averaged polar magnetic field, \cite{PSc18} 
 computed a solar dynamo amplitude (SODA) index. They used the SODA index as a precursor 
 for predicting the next cycle's amplitude and estimated a maximum SSN V2.0 of 135$\pm$25 
 for cycle 25. The Fe XIV coronal green line emission appears at high latitudes 
 ($\ge$ $50^{\circ}$) just before solar cycle maximum, which subsequently drifts to the 
 poles. This is known as rush-to-the-poles (RTTP). Based on the correlations of the rise 
 rate of the RTTP to the delay time between the end of the RTTP and the maximum of the 
 following cycle, \cite{PeN18} estimated the maximum SSN V2.0 of cycle 25 of 130. Similarly, 
 using the surface flux transport model \cite{JiW18} predict the polar field strength at 
 the end of cycle 24 and predicted the amplitude of cycle 25 to be 125$\pm$32, indicating 
 a 10\% stronger cycle 25 than cycle 24. Also, using continuous century-scale data-driven 
 surface flux transport simulations, \cite{BNa18} reported a slightly stronger solar cycle 
 25 than cycle 24 with an SSN${_{max}}$ ranging between 109 and 139 (V2.0). Thus, 
 using the revised SSN V2.0 and the other existing correlations that relate to the strength 
 of the following cycle, the other researchers also arrived at the same conclusion for 
 the amplitude of cycle 25 similar to the relatively stronger cycle 25 as proposed by this 
 study. The relatively stronger upcoming cycle 25 can be understood from the fact that the 
 axial dipole moment during cycle 24 (by Dec. 2017) has been stronger than that 
 during cycle 23 \citep{JiW18}. This is in keeping with the flux transport dynamo model by 
 \cite{CCJ07} wherein, the authors predicted a weaker cycle 24 based on the weaker axial 
 dipole moment during cycle 23.  
 With space missions like the Parker Solar Probe being operational and upcoming 
 solar missions like the ADITYA-L1 mission, by India, planned for launch in 
 2020 \citep{JaS17}, cycle 25 is bound to reveal more crucial insights 
 into as yet unknown aspects of the internal workings of our sun.

\acknowledgments
This work has made use of NASA's OMNIWEB services Data System. 
 The authors thank the free data use policy of the National Solar
 Observatory (NSO/KP, NSO/SOLIS and NSO/GONG), OMNI2 from NASA 
 and WDC-SILSO at Royal Observatory, Belgium, Brussels. SKB 
 acknowledges the support by the PIFI (Project No. 2015PM066) 
 program of the Chinese Academy of Sciences and the NSFC 
 (Grant No. 11750110422, 11433006, 11790301, and 11790305). 
 SA acknowledges an INSA Honorary Scientist position.
%% ------------------------------------------------------------------------ %%
%% References and Citations

%%%%%%%%%%%%%%%%%%%%%%%%%%%%%%%%%%%%%%%%%%%%%%%
% BibTeX is preferred:
%
\bibliography{bibliography}
% don't specify bibliographystyle
%%%%%%%%%%%%%%%%%%%%%%%%%%%%%%%%%%%%%%%%%%%%%%%

% Please use ONLY \citet and \citep for reference citations.
% DO NOT use other cite commands (e.g., \cite, \citeyear, \nocite, \citealp, etc.).
%% Example \citet and \citep:
%  ...as shown by \citet{Boug10}, \citet{Buiz07}, \citet{Fra10},
%  \citet{Ghel00}, and \citet{Leit74}.

%  ...as shown by \citep{Boug10}, \citep{Buiz07}, \citep{Fra10},
%  \citep{Ghel00, Leit74}.

%  ...has been shown \citep [e.g.,][]{Boug10,Buiz07,Fra10}.

\end{document}